\documentstyle[12pt,preprint,aps,epsfig,floats, amssymb]{revtex}
\newcommand{\be}{\begin{equation}}
\newcommand{\ee}{\end{equation}}
\newcommand{\bea}{\begin{eqnarray}}
\newcommand{\eea}{\end{eqnarray}}
\newcommand{\ba}{\begin{array}}
\newcommand{\ea}{\end{array}}
\newcommand{\lsim}
{{\;\raise0.3ex\hbox{$<$\kern-0.75em\raise-1.1ex\hbox{$\sim$}}\;}}
\newcommand{\gsim}
{{\;\raise0.3ex\hbox{$>$\kern-0.75em\raise-1.1ex\hbox{$\sim$}}\;}}
\def\msQ{m^2_{\tilde Q_L}}
\def\msuL{m^2_{\tilde u_L}}
\def\msdL{m^2_{\tilde d_L}}
\def\msuR{m^2_{\tilde u_R}}
\def\msdR{m^2_{\tilde d_R}}
\def\mslL{m^2_{\tilde L_L}}
\def\mseL{m^2_{\tilde e_L}}
\def\msnL{m^2_{\tilde{\nu}_L}}
\def\mseR{m^2_{\tilde e_R}}
\def\mhU{m^2_{H_u}}
\def\mhD{m^2_{H_d}}

\def\st{\sin^2 \theta_W}

\tighten
\begin{document}
\preprint{
\noindent
\begin{minipage}[t]{2in}
\begin{flushleft}
\end{flushleft}
\end{minipage}
\hfill
\begin{minipage}[t]{2in}
\begin{flushright}
September 2003\\
JLAB-PHY-03-133 \\
IISc-CTS-4/03\\
\tt{hep-ph/0312361}\\
\vspace*{.2in}
\end{flushright}
\end{minipage}
}
\draft
\title{\bf  Squark and slepton
masses as probes of supersymmetric $SO(10)$ unification}
\bigskip
\author{B. Ananthanarayan}
\address{Thomas Jefferson National Acceleratory Facility,
Newport News, Virginia 23606, USA \\
Centre for Theoretical Studies, Indian Institute of 
Scince, Bangalore, India\footnote{Permanent address}} 
\vspace{2mm}
\author{P. N. Pandita}
\address{Department of Physics,
North-Eastern Hill University,  Shillong 793 022, India}

\maketitle

\begin{abstract}
We carry out an analysis of the 
non-universal supersymmetry breaking scalar 
masses arising in $SO(10)$ supersymmetric unification.
By considering patterns of squark and slepton masses, 
we derive a set of sum rules for the sfermion masses  
which are independent of the manner in which $SO(10)$ breaks 
to the Standard Model gauge group via its
$SU(5)$ subgroups.
The phenomenology arising from such non-universality
is unaffected by the symmetry breaking pattern, so long as 
the breaking occurs via any of the  $SU(5)$ subgroups of the $SO(10)$ group. 
\end{abstract}

\vspace{3mm}

{\it Keywords:  Supersymmetry,  unification,
renormalization group analysis, sum rules}
\pacs{PACS number(s):  12.10.Dm, 12.10.Kt, 14.80.Ly}

\section{Introduction}
Grand unification~\cite{Ross} of the standard model gauge group into 
a simple group with the matter and Higgs particles transforming
under irreducible representations of this simple
group is an attractive framework for pursuing physics beyond
standard model~(SM). 
On the other hand supersymmetry~\cite{wess} is at present the only 
framework in which the Higgs sector of the Standard Model~(SM), 
so crucial for its internal consistency, is natural.  
A much favored implementation of
the idea of low energy supersymmetry~(SUSY) is the Minimal
Supersymmetric Standard Model~(MSSM)~\cite{Nilles}, which is obtained by 
doubling the number of states of the SM, and introducing
a second Higgs doublet (with the opposite hypercharge of the
SM Higgs doublet) to generate masses for all the fermions
and to cancel triangle gauge anomalies. The MSSM leads to a successful 
prediction~\cite{Amaldi} for the ratios of the three
gauge couplings, with a unification scale of $M_G \simeq 10^{16}$ GeV 
and supersymmetric thresholds near the electroweak scale ($M_S\simeq
1$ TeV).  
The possibility of such a unification scale in the supersymmetric context
would naturally suggest the framework of supersymmetric grand unification 
at the scale $M_G$. Furthermore, 
the presence of supersymmetry would also make the hierarchy
between the disparate scales  associated with the grand unification~($M_G$),
and the weak scale $M_W \sim 100$ GeV technically natural.
However, despite many years of dedicated experimental work, no direct
evidence has been found to vindicate 
the idea of unification, especially its most heralded prediction of 
proton decay.  All the present experimental data is, however, consistent with
the idea of low energy supersymmetry  with a grand unified scale being
around $10^{16}$ GeV.  It is therefore important to keep
an open mind toward all possibilities, conventional and
unconventional, in our search for unification of strong and electroweak 
interactions, since there is going to be a concerted effort
to search for collider signatures of such possibilities
at the next generation of experiments~\cite{Freitas}.

At the theoretical level, the  gauge group $SU(3)\times
SU(2)\times U(1)$  remains a completely unexplained feature of 
the standard model.
However, it can be elegantly unified into simple groups like
$SU(5)$~\cite{GG1} or $SO(10)$~\cite{SO10}.  The $SO(10)$ group is
particularly appealing,  since it has a ${\bf 16}$
dimensional representation which is large enough to accomodate
an entire generation of standard model fermions, and can further
accomodate a right-handed neutrino as well.  
The $SO(10)$ framework can readily be extended to include  supersymmetry.  
Furthermore,  a complex
${\bf 10}$ dimensional representation of $SO(10)$ can be employed
to accomodate the two Higgs doublets of the low energy minimal 
supersymmetric standard model.    

Unlike the case of $SU(5)$, the rank of the group
$SO(10)$ exceeds that of the standard model by 1. 
The breaking of the additional $U(1)$ factor
in $SO(10)$ group leads to  characteristic
non-universal contributions to the otherwise
universal soft SUSY breaking scalar masses of the minimal
supersymmetric models.  These additional corrections
to the soft scalar masses are usually referred to as $D-$term 
contributions, and lead to non-universality for the 
soft scalar masses . In general $D-$term contributions
to the SUSY breaking soft scalar masses arise whenever a gauge symmetry is 
spontanesously broken with a reduction of rank~\cite{Drees}. 
These $D-$term contributions have important phenomenological 
consequences at low energies
as they allow one to reach certain regions of parameter space which are 
not otherwise accessible with  universal boundary 
conditions~\cite{KM1,Autoetal,BKP1}.
Non-universality, and in particular
the $D-$term contributions,  may have a dramatic 
impact on the sum rules~\cite{MR1}
satisfied by the squark and slepton masses. Such effects are likely
to help distinguish between different scenarios for breaking of grand unified
symmetry at high energies~\cite{KMYPLB,KMYPRD,CH1}.

In this work we systematically consider the $D-$ term non-universality
that is generated in $SO(10)$ unification.  
This is the simplest of grand unification models  where
a $D-$ term non-universality is inevitable.  Indeed,
since $SO(10)$ contains $SU(5)\times U(1)$ as one its ``natural''
subgroups, one common route for breaking $SO(10)$ is
via $SO(10)\to SU(5)\times U(1)_Z$, with $SU(5)$ subsequently
breaking via $SU(5) \to SU(3)_C\times SU(2)_L \times U(1)_X$. 
However, in the breaking of $SO(10)$ via $SU(5)$, there exist 
two different ways of embedding $SU(3)_C\times SU(2)_L \times U(1)_Y$ into
$SU(5)\times U(1)$. In the ``conventional'' embedding~\cite{SO10} via $SU(5)$, 
the hypercharge generator $Y$ is identified with the generator 
$X$ of $U(1)_X$. On the other hand, in the 
``flipped''~\cite{DRGG,SMB1,SMB2} embedding
the hypercharge generator is identified with a linear combination of
the generators $X$ and $Z$. We shall study in detail the consequences
of these two different embeddings  of the SM gauge group in $SO(10)$.
We should mention here that apart from the ``natural'' subgroup 
$SU(5)\times U(1)$, the group $SO(10)$ also has ``natural'' 
subgroup $SO(6)\times SO(4).$
Since $SO(6)$ is isomorphic to $SU(4)$ and $SO(4)$  is isomorphic
to $SU(2) \times SU(2)$, $SO(10)$ contains the group
$SU(4)\times SU(2) \times SU(2)$. We shall not consider this pattern of
breaking of $SO(10)$ in this paper.

The plan of this paper is as follows. In Sec. II, we consider in 
detail the the $D-$term non-universality in supersymmetric $SO(10)$ 
unification. We shall review the embedding of the standard model gauge group
in the $SO(10)$ group.  We shall identify the hypercharge generator $Y$
for conventional and flipped embeddings, and the orthogonal generator
$Y^\perp$ whose eigenvalues determines the $D-$ term contributions
at the unification scale.
In Sec. III, we write down the renormalization group
equations for the sfermion masses and their solutions, and 
derive  sum rules for these masses in generality for the
embeddings of the SM gauge group in the $SO(10)$ group.
For the conventional embedding of $SO(10)$, it turns
out that our sum rules are same as those that were discovered 
for $SU(5)$ unification  with non-universal masses~\cite{CH1}.
We trace the origin of this to the fact that the $D$-term non-universality
in the conventional embedding of $SO(10)$ arises in such a manner that it 
is degenerate within the ${\bf 10}$ and ${\bf 5}^*$ representations of 
$SU(5)$.  On the other hand we find the  result that the 
sparticle sum rules for flipped embedding, 
where the $D-$ term non-universality is not expected to have a simple 
pattern, are identical to the case of conventional embedding.
By considering the nature of the embedding of the conventional and
flipped scenarios is some detail we establish why this is the case.
A further surprising aspect we find  is that 
the individual masses entering the sum rules  also do not
depend on the manner in which the SM is embedded into $SO(10)$.

In Sec. IV, we present results for the sparticle spectrum 
based on numerical studies and our conclusions. 
We note that since $D$ term contributions to the scalar masses at the
unification scale~\cite{Autoetal} has received considerable attention,
our results show that it is immaterial whether the embedding is
the conventional or the flipped one.
One input we have employed is the requirement that
the fine tuning problem inherent in the embedding of 
MSSM in $SO(10)$~\cite{BFT1} 
be  alleviated, in both the conventional and flipped
scenarios, by the $D$ term contribution, thereby enabling us to
fix the sign of this contribution. 
To our knowledge, this is the first time that this has been
pointed out in the context of flipped unification. 
Noting that the fine tuning problem is further alleviated by
the effects of the right-handed neutrino Yukawa coupling to the
renormalization group flow from the unification scale $M_G$ to
a scale $M_R\sim 10^{14.5}$ GeV, a scale  inspired by 
solar neutrino physics~\cite{BPZ1}, it is possible to 
render supersymmetric $SO(10)$ unification into a technically natural 
and viable candidate for a complete theory, with unique signatures.

\bigskip

\section{Conventional and Flipped Embedding of the SM in $SO(10)$ }
We begin by noting a crucial difference between the $SO(10)$ 
GUT model and the $SU(5)$
GUT model. Whereas the embedding of the SM gauge group 
into a unified gauge group is unique for $SU(5)$,  
there are two different ways in which the hypercharge group can be 
embedded in $SO(10)$.
These different embeddings 
may lead to different predictions for the low energy physics.

In  $SO(10)$ grand unification, all the matter particles of one family
of the Standard Model~(SM) together with a right handed neutrino belong
to the spinor representation ${\bf 16}$.  Each such spinor representation
${\bf 16}$ can be decomposed under the maximal subgroup
$SO(10) \supset SU(5) \times U(1)_Z$ as
\bea
{\bf 16} = {\bf 5^*}_{3} + {\bf 10}_{-1} + {\bf 1}_{-5}, \\
{\bf 10} = {\bf 5}_2 +{5^*}_{-2}.
\eea
Furthermore under $SU(5) \supset SU(3)_C \times SU(2)_L \times U(1)_X,$ 
we have the decomposition
\bea
{\bf 5}& = & ({\bf 3},{\bf 1})_{-2} + ({\bf 1},{\bf 2})_{3} , \nonumber \\
{\bf 5^*}& = & ({\bf 3^*},{\bf 1})_2 + ({\bf 1},{\bf 2})_{-3} ,
\nonumber \\
{\bf 10} & = & ({\bf 3},{\bf 2})_1 + ({\bf 3^*},{\bf 1})_{-4}
+({\bf 1},{\bf 1})_6 , \nonumber \\
{\bf 1}  & = & ({\bf 1},{\bf 1})_0.
\eea
We note that $U(1)_X$, which is the subgroup of $SU(5)$, is not identical
with the $U(1)_Y$  of the SM at this stage. Note also that each 
${\bf16}$ includes two pairs of $({\bf 3^*},{\bf 1})$ and  $({\bf 1},{\bf 1}).$ 

In order to identify the hypercharge group, we consider the 
decomposition $SO(10) \supset SU(5)\times U(1)_Z \supset
SU(3)_C \times SU(2)_L \times U(1)_X \times U(1)_Z$. 
Therefore, the hypercharge $U(1)_Y$ must be a linear combination of 
$U(1)_X$ and $U(1)_Z$, i.e. $U(1)_Y \subset U(1)_X \times U(1)_Z$.
Thus, there are two ways to define the  hypercharge generator of the 
SM:
\begin{enumerate}
\item $Y = X,$ 
\item $Y = -\frac{1}{5}(X+6Z),$
\end{enumerate}
upto an overall normalization factor.
The first case corresponds to the the Georgi-Glashow model~\cite{GG1},
whereas the second identification of hypercharge corresponds to the flipped 
case~\cite{DRGG,SMB1}.

In the first case the $U(1)$ generator of $SO(10)$ that is orthogonal to
$Y$ and the diagonal generators of $SU(3)_C$  and $SU(2)_L$ is 
\bea
Y^{\bot} &=& -Z, 
\eea
whereas in the flipped case
we have for the orthogonal generator\footnote{
In order to determine this, one must work with a set of
normalized generators.  The explicit normalization factors 
are adapted from the expressions given in ref.~\cite{SMB1}.}
\bea
Y^{\bot} &=& {-4X+Z\over 5},
\eea
We will comment on the normalization in the following.
After a suitable identification of the fields lying in the relevant 
representations of $SO(10)$, the effect of $SO(10)$ breaking at the unification
scale leads to $D$-term non-universality, which is computed in terms of
the eigenvalues of the operator $Y^\perp$ on the fields  
which will discuss in the next section.

\section{Solutions of Renormalization Group Equations
and sum rules}

For the squarks and sleptons of the
first and second family(light generations), 
the renormalization group~(RG) equations for the soft scalar 
masses are given by 
\bea
16 \pi^2 {d \msQ\over dt} &=&
-{32\over 3} g_3^2 M_3^2 - 6 g_2^2 M_2^2 - {2\over 15} g_1^2 M_1^2
+ {1\over 5} g_1^2 S, 
\label{rg1}
\\
16 \pi^2 {d \msuR \over dt} &=&
-{32\over 3} g_3^2 M_3^2 - {32\over 15} g_1^2 M_1^2
- {4\over 5} g_1^2 S, 
\label{rg2}\\
16 \pi^2 {d \msdR \over dt} &=&
-{32\over 3} g_3^2 M_3^2 - {8\over 15} g_1^2 M_1^2
+ {2\over 5} g_1^2 S, 
\label{rg3}\\
16 \pi^2 {d \mslL \over dt} &=&
- 6 g_2^2 M_2^2 - {6\over 5} g_1^2 M_1^2
- {3\over 5} g_1^2 S, 
\label{rg4}\\
16 \pi^2 {d \mseR\over dt} &=&
- {24\over 5} g_1^2 M_1^2
+ {6\over 5} g_1^2 S, 
\label{rg5}
\eea
where $t \equiv {\rm ln}(Q/Q_0)$, with $Q_0$ being some initial large
scale; $M_{3,2,1}$ are the running gaugino masses, $g_{3,2,1}$ are the 
usual gauge couplings associated with the SM gauge group, and 
\be
S \equiv {\rm Tr}(Ym^2) = m_{H_u}^2 - m_{H_d}^2 + \sum_{\rm families}
(\msQ - 2 \msuR + \msdR - \mslL + \mseR)\> .
\ee
The $U(1)_Y$ gauge coupling $g_1$ (and $\alpha_1$)  is taken to be in
a GUT normalization throughout this paper. The quantity S evolves according to
\be
{dS \over dt} = {66\over 5} {\alpha_1\over 4 \pi}S
\ee
which has the solution
\be
S(t) = S(t_G) {\alpha_1(t) \over \alpha_1(t_G)}.
\ee
We note that if $S=0$ at the initial scale, which would be the case if
all the soft sfermion and Higgs masses are same, then the RG evolution 
will maintain it to be zero at all scales.  

The solution for the 
renormalization group equations (\ref{rg1})--(\ref{rg5})  can then  
be written as
\bea
\msuL &=& \msQ(t_G) +  C_3 + C_2 + {1\over 36} C_1
+ ({1\over 2} - {2\over 3} \st) M_Z^2 \cos (2 \beta) + {1\over 5} K,
\\
\msdL &=& \msQ(t_G) +  C_3 + C_2  + {1\over 36}  C_1
+ (-{1\over 2} +{1\over 3} \st) M_Z^2 \cos (2 \beta)+ {1\over 5} K,
\\
\msuR &=& \msuR(t_G) + C_3                  + {4\over 9}  C_1
+ {2\over 3} \st M_Z^2 \cos (2 \beta) - {4\over 5} K,
\\
\msdR & =& \msdR(t_G) +  C_3                 + {1\over 9} C_1
- {1\over 3} \st M_Z^2 \cos (2 \beta)+ {2\over 5} K,
\\
\mseL & =& \mslL(t_G)                + C_2 + {1\over 4}   C_1
+ (-{1\over 2} + \st)  M_Z^2 \cos (2 \beta)  - {3\over 5} K,
\\
\msnL & = & \mslL(t_G)               +  C_2 + {1\over 4}  C_1
+ {1\over 2} M_Z^2 \cos (2 \beta) - {3\over 5} K,
\\
\mseR & =& \mseR(t_G)                            +   C_1
-\st M_Z^2 \cos (2 \beta) + {6\over 5} K,
\eea
where $C_1,\, C_2$ and $C_3$ are given by
\bea
C_i(t)= {a_i\over 2 \pi^2} \int_t^{t_G} g_i(t)^2 M_i(t)^2, \, i=1,2,3\\
a_1={3\over 5}, a_2={3\over 4}, a_3={4\over 3},
\eea
and
\bea
K = \frac{1}{16\pi^2}\int^{t_G}_t g_1^2(t)~S(t)~dt
  = \frac{1}{2b_1}S(t)\left[1 - \frac{\alpha_1(t_G)}{\alpha_1(t)}\right],
\eea
is the contribution of the non-universality parameter $S$ to the 
sfermion masses, and $b_1=-33/5$. 

We now proceed to determine the combinations of sfermion masses
that would satisfy sum rules which are independent 
of the scalar masses at $t_G$, 
and the  $D-$ term contribution. We first consider the conventional case. 
To this end,  we note that the 
right hand side of the equations given above that determine the sfermion masses
contain a term proportional to $Y_i~K$, which is  proportional
to $Y_i ({\rm Tr} Y m^2)$,and the index $i$ runs over all the sfermions,
$i=\tilde{Q}_L,\tilde{L}_L,\tilde{u}_R,\tilde{d}_R,\tilde{e}_R$.  
We are interested in finding those
combinations of squared sfermion masses where this contribution proportional to
$K$ vanishes.  Defining this combination as $\sum_i \kappa_i m_i^2$, 
where the constants $\kappa_i$ are to be determined, and
we have the condition
\begin{equation} \label{con1}
\sum_i \kappa_i Y_i=0.  
\end{equation}
We now recall  the  boundary
condition at the GUT scale which reads $m_i^2=m_0^2 + Y^\perp_i
g_{10}^2 D$, where $m_0$ denotes a generic universal soft mass
parameter for the sfermion.  For
the combination of squark and slepton masses where no unknown parameters enter, we require that
the universal mass term contribution as well as the D term
contribution vanish.  These  conditions translate into the equations
\begin{eqnarray}
& \displaystyle \sum_i \kappa_i=0, & \label{con2} \\
& \displaystyle \sum_i \kappa_i Y^\perp_i=0. & \label{con3}
\end{eqnarray} 
We can solve these three constraint equations for the five mass parameters,
and in general there is a family of solutions.  We arbitrarily fix
the coefficient of the right handed selectron in the sum rules 
to be $-1$, and ask for integer
values for the parameter $\kappa_{\tilde{Q}}$
for the squark doublet.
We then  arrive at the following two linearly independent sum rules:
\bea
2 m_{\tilde{Q}}^2 - \msuR -\mseR &=& (C_3 + 2C_2 - {25\over 18}C_1),
\label{sum1}\\
m_{\tilde{Q}}^2 + \msdR - \mseR - m_{\tilde{L}}^2
&=& (2C_3 - {10\over 9} C_1)\label{sum2},
\eea
where we have used the notation
\bea
m_{\tilde{Q}}^2={1\over 2} (\msuL + \msdL),  \, \,  \, m_{\tilde{L}}^2
={1\over 2} (\mseL + \msnL). \nonumber
\eea
The essential point is that we have for the conventional embedding 
$Y=X$ and $Y^\perp=-Z$,  and have solved the constraint equations  
(\ref{con1}) and (\ref{con3}).  Furthermore, 
the solutions we obtain for flipped embedding is exactly the same.
This follows from the fact that the flipped embeddding is a special
case of the solutions of the constraint equations for the general
case  $Y=a X + b Z$ and $Y^\perp=c X + d Z$, where $a,\, b,\, c,\, d$
are arbitrary constants.  This is not entirely unexpected, since in 
each of  the
$SU(5)$ symmetry breaking chains the $D$-term contribution
is the same, as each depends only on the gauge contribution.
This is one of the main results of this paper.
It explicitly demonstrates that despite the possibility that the model loses
its uniqueness in the choice of embedding, the gauge symmetry guarantees
that certain relations are preserved.  

The two sum rules expressed above, which are valid only when the 
symmetry breaking takes place via $SU(5)$ or flipped $SU(5)$,
denote two linearly independent sum rules with integer coefficients.
The particular choice we have presented is similar to the one
in~\cite{CH1}.
We have chosen to present our sum rules in this manner in order
to bring out this feature.  The reason for 
our ability to reproduce those sum rules 
is that the generic non-universality
that was considered in~\cite{CH1} for the masses of the 
${\bf 10}$ and ${\bf 5^*}$
has the same structure as is generated by  $D$- terms in $SO(10)$
unification in the conventional case, where the non-universality
is now explicitly given by considering the eigenvalues
of the corresponding $Y^\perp$:  
\bea
\msQ(t_G) &=& \msuR(t_G) = \mseR(t_G) = m_{16}^2+g_{10}^2D,\\
\mslL(t_G) &=& \msdR(t_G) = m_{16}^2-3g_{10}^2D,\\
\mhU(t_G) &=& m_{10}^2-2g_{10}^2D,\\
\mhD(t_G) &=& m_{10}^2+2g_{10}^2D,
\eea
at the $SO(10)$ unification scale $M_G$. Here $m_{16}$ and  $m_{10}$ 
are the common soft scalar masses, corresponding to the $\bf{16}$
and $\bf{10}$ dimensional representations, respectively of $SO(10)$, 
at the unification scale. 
We note here that in the breaking of
$SO(10)$ the rank is reduced by one, and hence the $D$-term contribution
to the soft masses is expressed by a single parameter $D$.
We note from the above that $S(t_G)=-4 g_{10}^2 D$.  
The solution for $K$ is obtained by eliminating 
$C_1$, $C_2$, $C_3$, $m_{16}^2$, and $m_{10}^2$  from the sfermion
mass equations. We get for the conventional embedding the result~\cite{CH1}:
\bea
K &=& {1\over 4}
(m_{\tilde{Q}}^2 - 2m_{\tilde{U}^c}^2 + m_{\tilde{D}^c}^2 +
m_{\tilde{E}^c}^2 - m_{\tilde{L}}^2 + {10\over 3}
\sin^2 \theta_W M_Z^2 \cos 2\beta)\label{conv}. 
\eea
\bigskip

We now consider the eigenvalues corresponding
to $Y^\perp$ for the flipped embedding.  We find that the result
is identical to that for the conventional embedding, as may be
explicitly checked.  Indeed, the normalization for $Y^\perp$ is chosen so as
to obtain this result.  The conclusion we draw is that
the gauge symmetry of $SO(10)$~(i) protects the sum rules 
irrespective of the embedding; (ii) 
the explicit $D-$ term contributions to the sfermion masses
for the two embeddings are in fact identical.  
We note that (ii) implies (i), but not the other way round.  
The stronger  result (ii)
is of significance to the phenomenology of $SO(10)$ models with
$D-$ term corrections, which is the subject of the next section.

\section{Sparticle spectrum and Conclusions}
In order to study the implications of the $D$- term non-universality
for the sparticle spectrum,
we have carried out numerical integration of the RG equations.
We integrate the system of coupled renormalization group equations
of the gauge couplings, Yukawa couplings, the gaugino masses, the
scalar mass squared parameters and the soft trilinear couplings of
the MSSM.  We employ well-known results for the unification scale,
the value of the unified gauge coupling constant and the unified
Yukawa coupling for the heaviest generation.  
We present the results of numerical analysis
that take into account the effects of the
$D-$ term non-universality on the boundary conditions for
scalar mass squared parameters in the conventional and flipped 
scenarios. For all other parameters
we assume universal boundary conditions.
In other words, 
starting with values of the common gaugino mass ($M_{1/2}$),
the scalar masses at the unification scale (parameterized by $m_0
(=m_{16}=m_{10})$ 
and $g_{10}^2 D$), trilinear couplings
($A$), and with the Yukawa couplings
having a unified value $h$ at the unification scale, 
and $\alpha_G$,  we  integrate
the set of coupled differential equations down to the effective
supersymmetry scale of $\sim 1$ TeV.  
Of the two minimization conditions for the Higgs potential, we recall that
one can be written as 
\begin{equation}
{\mu_1^2 -  \mu_2^2  \tan^2 \beta \over \tan^2\beta -1}= {m_Z^2\over 2}.
\end{equation}
Proceeding in the well-known fashion~\cite{AS1} of determining 
$\tan\beta$ from
the accurately known value of the $\tau$-lepton mass,  
and inserting it into the equation above, and using the values of
the Higgs mass squared parameters determined from the evolution of
the parameters yields,
from the relations
$\mu_1^2=m_{H_d}^2 + \mu^2$ and $\mu_2^2=m_{H_u}^2 + \mu^2$,
the parameter $\mu$.

We recall that for the universal case
sufficiently large values
of the common gaugino mass $M_{1/2}$ are required to ensure that 
the gluino is sufficiently heavy, 
and also fairly large values of $m_0$ ($< M_{1/2}$)
are required to ensure that the neutralino
is the LSP (in order to prevent the lightest slepton from becoming the
LSP). An upper bound on $m_0$ ensues when we  require  
a sufficiently large $m_A$ ($m_A=\mu_1^2+\mu_2^2$).  
Keeping these features in mind, we study the effects of the $D-$term
on the spectrum. 

In our work we have chosen the sign of $D$ to be positive.  This
alleviates the problem of fine-tuning, 
inherent in  $SO(10)$ unification, by
allowing $m_{H_u}^2$ to evolve to values that are negative and
larger in magnitude, compared to when the term is absent.
(We also note here that once solar neutrino parameters are better
determined, one would have to include the effects of the Yukawa 
neutrino coupling between $M_G$ and a scale $M_R\sim 10^{14.5}$ GeV
which would also have an effect of alleviating the fine-tuning
problem.)  
We begin with a simple illustration
for choices of the parameters of the model that would guarantee
radiative electroweak symmetry breaking, that the lightest
neutralino is the LSP and that $m_A$ is significantly larger than
$m_Z$.  In Figs.1 and 2 we plot the evolution of the
five sfermion mass squared parameters in a manner which is 
analogous to~\cite{KMYPLB}.
In Fig. 1(a) and 1(b) we illustrate the evolution of the mass parameters for
the case with $D=0$ for the two lightest generations
and the heaviest generation, respectively.
We have plotted the ratios $m_{\tilde{f}}^2/m_0^2$ vs. $t/{\rm Log}(10)$,
for all the sfermions of interest.
We have illustrated this for a typical
case of $M_{1/2}=800,\, m_0=700, A_0=0$ in units
of GeV and the common Yukawa coupling is taken to be $h_t=h_b=h_\tau=2.0$.
We now add a $D-$ term contribution with $g_{10}^2 D=300$ GeV, 
which is a fairly large contribution.
We illustrate this for the lighter generation (Fig. 2(a))
and for the heaviest generation (Fig. 2(b)). 
It is immediately apparent that for the right and left handed sleptons
of the heaviest generation
the renormalization group evolution  compensates for the 
initial mass splitting at the GUT scale. 
As seen clearly in Fig. 2(b), the near degeneracy of the 
two slepton states in the
absence of electroweak symmetry breaking makes the mixing between them,
once $SU(2)\times U(1)$ is broken, significant.  
Indeed, it is important
to observe the variation of the mass of the lighest slepton, since it has
the tendency to become lighter than the lightest neutralino and to emerge
as a candidate for the LSP, and is therefore not acceptable.  

We have also plotted in Fig. 3 curves corresponding to 
important observables.
We begin by observing the behaviour of $\mu$ which is seen to 
increase significantly  as $D'\equiv g_{10}^2 D$ is increased.
Note that the parameter $\mu$ is an important measurable quantity
in the chargino spectrum.  The pseudoscalar Higgs mass $m_A$
which is also plotted in the same figure is 
primarily responsible, apart from $\tan\beta$, 
for determining the properties of the entire Higgs spectrum of the model.
We have not plotted the value of the
mass of the lightest Higgs boson
since in the regime of $m_A\gg m_Z$ and
$\tan\beta \gg 1$, its value is a constant of about 120-125 GeV.
It has been pointed out that one can view the
rise of $m_A$ as a  relaxation of fine tuning of the parameters
in the model~\cite{CH1}.
In Fig. 3 the smaller mass eigenvalue resulting from the slepton mass 
squared mixing matrix  marked $\tilde{\tau}$
is plotted.  Indeed it rises for a while until the mixing effects become
pronounced and begins to fall.  In particular, the value of
$D'=$ 300 GeV is where the lighter slepton becomes degenerate
with the bino whose mass is $\sim$ 350 GeV when $M_{1/2}=800$ GeV.
We have also varied the parameters of the model and find the
trend to be maintained for reasonable variations.

To summarize, in this  work we have considered what is perhaps the simplest 
example of the $D-$ term non-universality following from $SO(10)$ unification.
We have obtained  sum rules for sfermion masses for
the conventional and flipped scenarios in a very general manner
and the method may be easily adopted for scenarios based on other
groups, e.g., $E_6$, etc.
It is interesting to note that the same sum rules emerge in
conventional and flipped embeddings.  
It turns out that the results are similar to that of
non-universality of the type discussed in refs.~\cite{KMYPLB,KMYPRD}.  
The presence of $D-$ term non-universality assists in rendering the
model to be technically natural.
\section{ Acknowledgements}  
BA would like to thank the 
Department of Science and Technology, Government of India,
whereas PNP would like to thank the
University Grants Commission,  for support during the course of this work.
PNP would also like to thank the Abdus salam International Centre
for Theoretical Physics, Trieste 
for its hospitality while part of this work was done. 
This work is supported by the Department of Energy under contract
DE-AC-05-84ER40150.


\newpage
\noindent{\bf Figure Captions}

\bigskip

\noindent {\bf Fig. 1 (a)} Evolution of the ratio
of the sfermion mass squared parameters to the universal mass squared
parameter, as a function of the momentum scale, for the lightest
generations. Here we have taken  $D=0$.  
The values of the other parameters are given in the text.

\medskip

\noindent {\bf Fig. 1 (b)} Evolution of the ratio
of the sfermion mass squared parameters to the universal mass squared
parameter, as a function of the momentum scale, for the heaviest
generation. Here we have taken  $D=0$.  
The values of the other parameters are given in the text.

\medskip

\noindent {\bf Fig. 2 (a)} Evolution of the ratio
of the sfermion mass squared parameters to the universal mass squared
parameter, as a function of the momentum scale, for the lightest
generations with $g_{10}^2 D=300$ GeV.
The values of the other parameters are given in the text.

\medskip

\noindent {\bf Fig. 2 (b)} Evolution of the ratio
of the sfermion mass squared parameters to the universal mass squared
parameter, as a function of the momentum scale, for the heaviest
generation with $g_{10}^2 D=300$ GeV. 
The values of the other parameters are given  in the text.

\medskip

\noindent {\bf Fig. 3 } Values of mass parameters at the supersymmetry
breaking scale, for the choice of parameters at the unification scale of
$M_{1/2}=800$, $m_0=700$, $A_0=0$, all in GeV,
and $h_t=h_b=h_\tau=2.0$, when $D'$ is varied from 0 to
400 GeV.

\begin{center}
\begin{figure}
\epsfig{figure=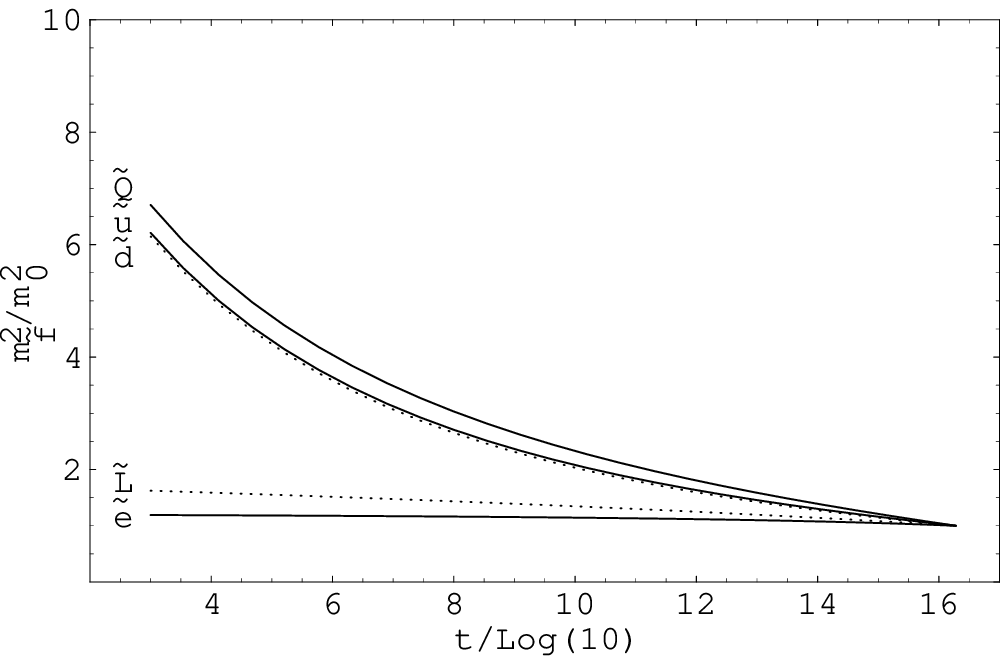,width=10cm,height=10cm}
\centerline{Fig. 1 (a)}
\epsfig{figure=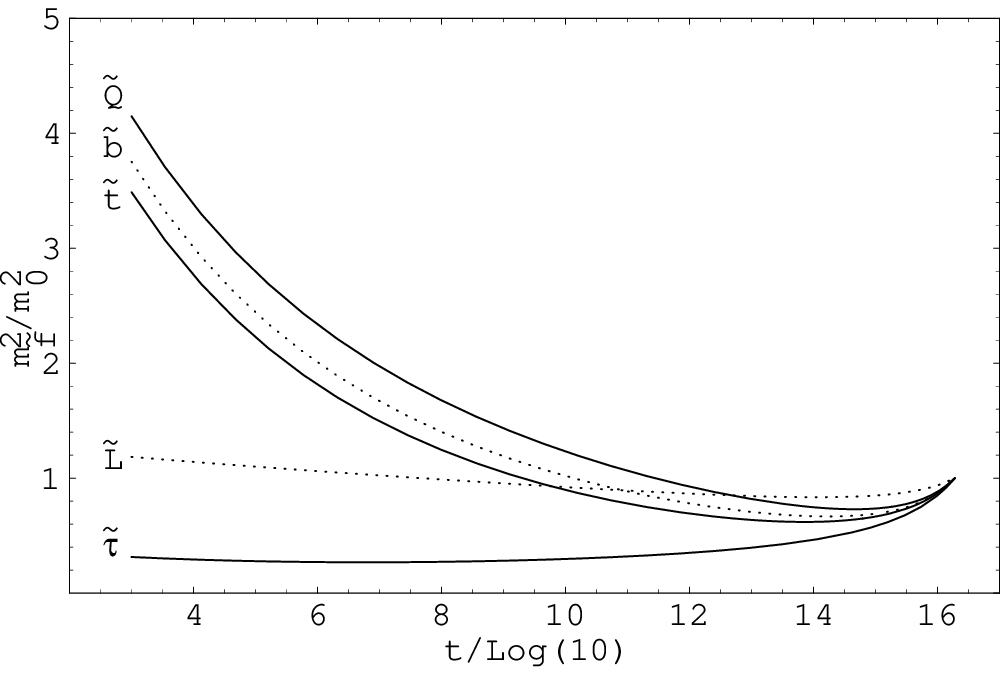,width=10cm,height=10cm}
\centerline{Fig. 1 (b)}
\end{figure}
\end{center}

\begin{center}
\begin{figure}
\epsfig{figure=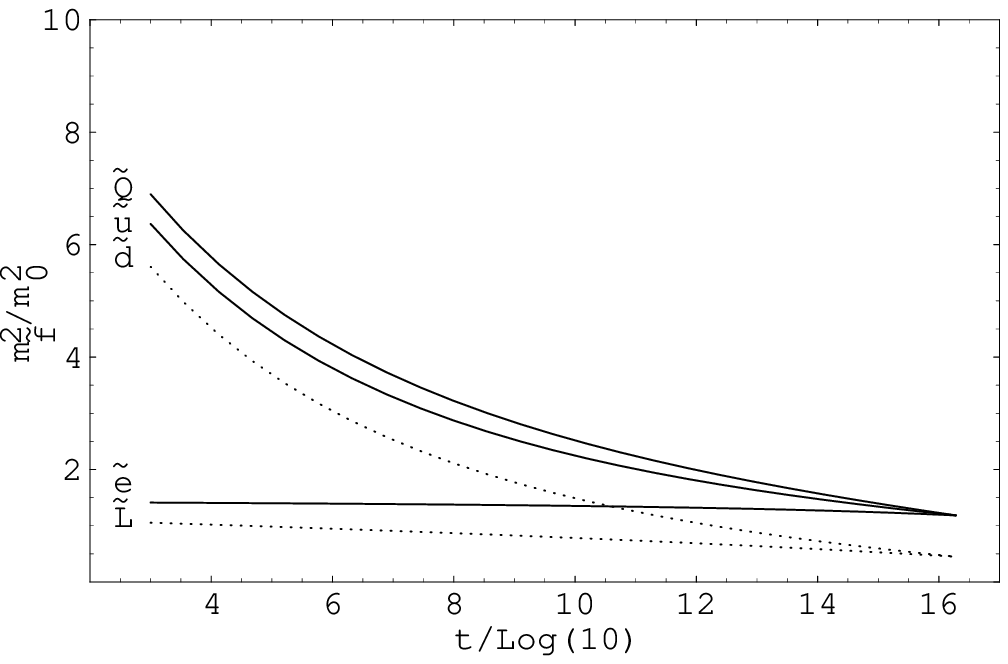,width=10cm,height=10cm}
\centerline{Fig. 2 (a)}
\epsfig{figure=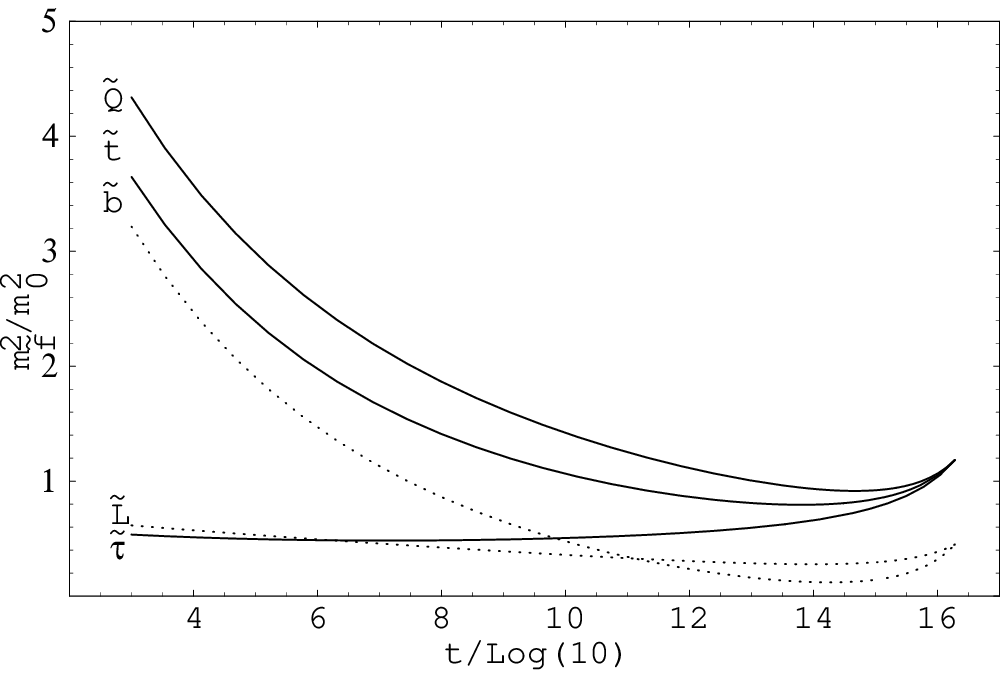,width=10cm,height=10cm}
\centerline{Fig. 2 (b)}
\end{figure}
\end{center}

\begin{center}
\begin{figure}
\epsfig{figure=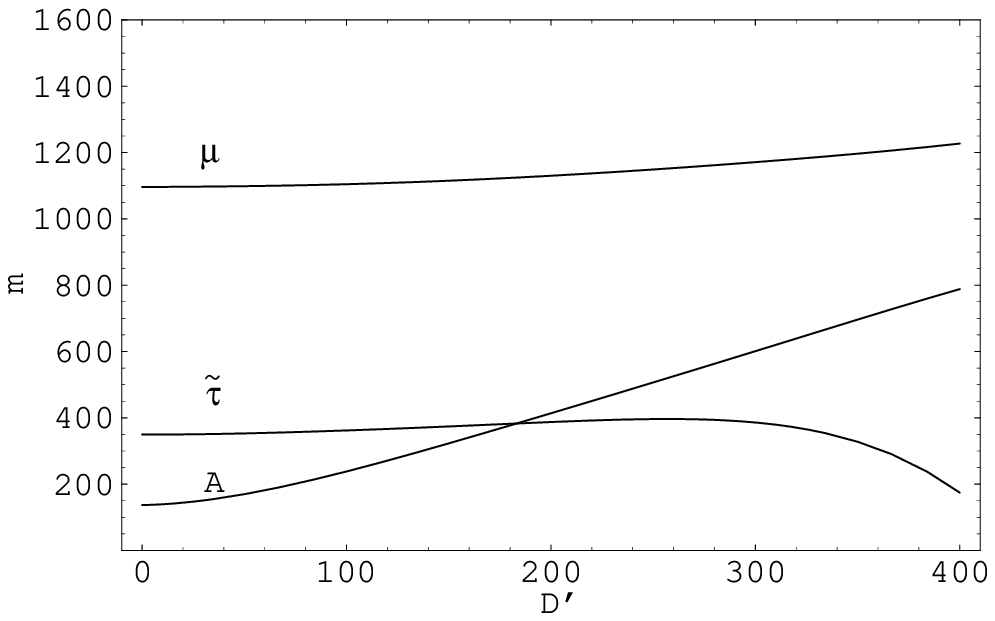,width=10cm,height=10cm}
\centerline{Fig. 3} 
\end{figure}
\end{center}

\begin{thebibliography}{abcdef}
\bibitem{Ross} G.~G~.Ross,
{\it Grand Unified Theories,} Benjamin/Cummings, 1984,
Reading, USA. 

\bibitem{wess} J. Wess and J. Bagger, ``Supersymmetry and Supergravity''
(Princeton University Press, Princeton, NJ, 1992);
P. Nath, R. Arnowitt and A. H. Chamseddine, `` Applied N = 1 Supergravity''
(World Scientific, Singapore, 1984).

\bibitem{Nilles}           
H.~P.~Nilles,
Phys.\ Rept.\  {\bf 110}, 1 (1984).


\bibitem{Amaldi}       
U.~Amaldi, W.~de Boer and H.~Furstenau,
Phys.\ Lett.\ B {\bf 260}, 447 (1991);
C.~Giunti, C.~W.~Kim and U.~W.~Lee,
Mod.\ Phys.\ Lett.\ A {\bf 6}, 1745 (1991);
J.~R.~Ellis, S.~Kelley and D.~V.~Nanopoulos,
Phys.\ Lett.\ B {\bf 260}, 131 (1991);
P.~Langacker and M.~X.~Luo,
Phys.\ Rev.\ D {\bf 44}, 817 (1991).

\bibitem{Freitas}           
A.~Freitas {\it et al.},
arXiv:hep-ph/0211108.

\bibitem{GG1}            
H.~Georgi and S.~L.~Glashow,
Phys.\ Rev.\ Lett.\  {\bf 32}, 438 (1974).

\bibitem{SO10}
H.~Georgi in Proceedings of the American Institute of Physics,
C.~E.~Carlson, ed., New York (1975), 575;
H.~Fritzsch and P.~Minkowski,
Annals Phys.\  {\bf 93}, 193 (1975).

\bibitem{Drees} M.~Drees,
Phys.\ Lett.\ B {\bf 181}, 279 (1986).

\bibitem{KM1}            
C.~F.~Kolda and S.~P.~Martin,
Phys.\ Rev.\ D {\bf 53}, 3871 (1996)
[arXiv:hep-ph/9503445].

\bibitem{Autoetal}    
D.~Auto, H.~Baer, C.~Balazs, A.~Belyaev, J.~Ferrandis and X.~Tata,
JHEP {\bf 0306}, 023 (2003)
[arXiv:hep-ph/0302155];
H.~Baer, M.~A.~Diaz, J.~Ferrandis and X.~Tata,
Phys.\ Rev.\ D {\bf 61}, 111701 (2000)
[arXiv:hep-ph/9907211].

\bibitem{BKP1}            
T.~Blazek, S.~F.~King and J.~K.~Parry,
JHEP {\bf 0305}, 016 (2003)
[arXiv:hep-ph/0303192].

\bibitem{MR1}              
S.~P.~Martin and P.~Ramond,
Phys.\ Rev.\ D {\bf 48}, 5365 (1993)
[arXiv:hep-ph/9306314].

\bibitem{KMYPLB}             
Y.~Kawamura, H.~Murayama and M.~Yamaguchi,
Phys.\ Lett.\ B {\bf 324}, 52 (1994)
[arXiv:hep-ph/9402254].

\bibitem{KMYPRD}              
Y.~Kawamura, H.~Murayama and M.~Yamaguchi,
Phys.\ Rev.\ D {\bf 51}, 1337 (1995)
[arXiv:hep-ph/9406245].


\bibitem{CH1}
H.~C.~Cheng and L.~J.~Hall,
Phys.\ Rev.\ D {\bf 51}, 5289 (1995)
[arXiv:hep-ph/9411276].

\bibitem{DRGG}
A.~De Rujula, H.~Georgi and S.~L.~Glashow,
Phys.\ Rev.\ Lett.\  {\bf 45}, 413 (1980).

\bibitem{SMB1}
S.~M.~Barr,
Phys.\ Lett.\ B {\bf 112}, 219 (1982).

\bibitem{SMB2} For a review, see, e.g.,      
S.~M.~Barr,
Phys.\ Rev.\ D {\bf 40}, 2457 (1989).


\bibitem{BFT1} For a recent discussion,  see, e.g.,              
H.~Baer, J.~Ferrandis and X.~Tata,
Phys.\ Lett.\ B {\bf 561}, 145 (2003)
[arXiv:hep-ph/0211418].


\bibitem{BPZ1}          
G.~A.~Blair, W.~Porod and P.~M.~Zerwas,
Eur.\ Phys.\ J.\ C {\bf 27}, 263 (2003)
[arXiv:hep-ph/0210058].





\bibitem{AS1} For a review, see, e.g., B. Ananthanarayan and Q. Shafi,
in ``Yukawa Couplings and Origin of Mass,'' P. Ramond ed., International
Press, 1996, Cambridge, USA, pp. 10;  see also
M.~E.~Gomez, G.~Lazarides and C.~Pallis,
Nucl.\ Phys.\ B {\bf 638}, 165 (2002)
[arXiv:hep-ph/0203131].

\end{thebibliography}
\end{document}